\documentclass[onecolumn]{elsart}

\usepackage{graphicx}

\newcommand{\meio}{\frac{1}{2}}

\begin{document}

\setlength{\parindent}{0.5cm}

\begin{frontmatter}

\title{Williams and Bjerknes model with growth limitation}

\author{S. C. Ferreira Junior\corauthref{cor}}
\corauth[cor]{Corresponding author.}
\ead{silviojr@fisica.ufmg.br}

\address{Departamento de F\'{\i}sica, Instituto de Ci\^encias Exatas,
Universidade Federal de Minas Gerais, CP 702, 30161-970, Belo Horizonte, MG,
Brazil}

\date{\today}

\begin{abstract}
Williams and Bjerknes proposed a simple stochastic  growth model to describe
the tumor growth in the basal layer of an epithelium. In this work we
generalize this model by including the possibility of saturation in the tumor
growth as it is clinically observed. The time evolution of the average number
of tumor cells and its variance for both the original and extended models are
studied by analytical methods and numerical simulations. The generated growth
patterns can be compact, connected or disconnected depending on the model
parameters used, and their geometrical properties are characterized through the
gyration radius, the number of interfacial cells and the density of empty sites
inside the patterns. \end{abstract}

\end{frontmatter}
\section{Introduction}
\label{sc1}
The population dynamics, including the growth of normal and tumor cells, is a
traditional problem investigated in Physics and Biology \cite{biomod}. In their
original work, Williams and Bjerknes \cite{wb}  built a model (WB model) to
describe the tumor growth in the basal layer of an epithelium. In their model,
one phase (the tumor cells) grows faster than the other (normal cells) by a
factor $\kappa$, which represents the carcinogenic advantage. This model
exhibits two distinct behaviors: unrestricted growth ($\kappa >1$) and complete
regression of the tumor ($\kappa\le 1$). In the special case $\kappa=1$ the
tumor always disappears due to the fluctuations. It is worthwhile to mention
that the WB model can also be applied to describe the growth of many systems
involving two competing phases. Using an adequate time step, the WB model can
be interpreted as a \lq {\it birth and death}\rq ~ process \cite{vankampen}
with constant division and death rates. Thus, all the well known results for
stochastic processes can be used to understand this model.

However, real tumors exhibit, in addition to the two distinct behaviors
previously  described, a quiescent state in which the tumor size remains
constant for a long time \cite {latency}. The underlying aspects of tumoral
biodynamic diversity involve a complex set of biochemical processes  and
environmental constraints  such as local nutrient availability, mechanical
stress,  immune response, etc. \cite{latency}. Numerous models of cancer growth
have been recently proposed in order to describe the tumor progression,
providing auxiliary tools for cancer diagnosis and therapy \cite{Chaplain}.

In this paper, we are proposing a generalization of the WB model in which  the
division and death rates of tumor cells depend on their total number.
Specifically, the probability of cell division decreases  whereas the cell
death probability increases as  the number of cancer cells increases. The paper
is organized as follows. In section \ref{sc2}, the original and the extended WB models
are defined. Sections \ref{sc3} and \ref{sc4} are dedicated to the analytical study of both
discrete and continuous time versions of these models. In section \ref{sc5}, the
geometrical properties of the growth patterns generated by the extended model
are characterized through computer simulations. Finally, some conclusions are
drawn in section \ref{sc6}.

\section{The extended Williams and Bjerknes models}
\label{sc2}

In the original WB model \cite{wb} the tissue is represented by a
two-dimensional  lattice where occupied sites represent tumor and empty sites
normal cells. All the sites are  initially empty,  except the center of the
lattice, since the tumor grows from a single malignant cell, in agreement  with
the theory of clonal origin of cancer \cite{Nowell}. The interfacial cells are
defined as those that have one or more nearest-neighbor sites of the opposite
type. The growth rule is very simple: one of the bonds between two opposite
cell types is chosen at random with equal probability; the normal cell of this
chosen bond is replaced by a tumor cell, with probability $g$, or  the tumor
cell is replaced by a normal one, with the complementary probability $r=1-g$.
In terms of the carcinogenic advantage $\kappa$, these probabilities can be
written as: \begin{equation} \label{probwb1} g=\frac{\kappa}{\kappa+1}
\end{equation}
and
\begin{equation}
\label{probwb2}
r=\frac{1}{\kappa+1}.
\end{equation}
The limit $\kappa=\infty$ corresponds to the Eden model \cite{eden},  more
specifically, the Eden model type B according to   the definitions given by
Vicsek \cite{vicsek}.

There are many variations of the WB model \cite{mf}, but we shall consider
only the cases that exclude the steps that do not change the pattern
configuration. Therefore, in all time steps the  number of tumor cells
increases or decreases by an unity with probabilities $g$ or $r$, respectively.
Particularly, we shall study the follwing variation of the original WB model:
at each time step a cell  type is chosen, either a tumor cell, with probability
$r$, or a normal cell, with probability $g$. Then, the selected cell is
converted to its opposite type. In other words, at each time step, either the
{\it division} or {\it death} of a single tumor cell occurs, with probabilities
$g$ or $r$, respectively.

\begin{figure}[h]
\begin{center}
\resizebox{8cm}{8cm}{\includegraphics{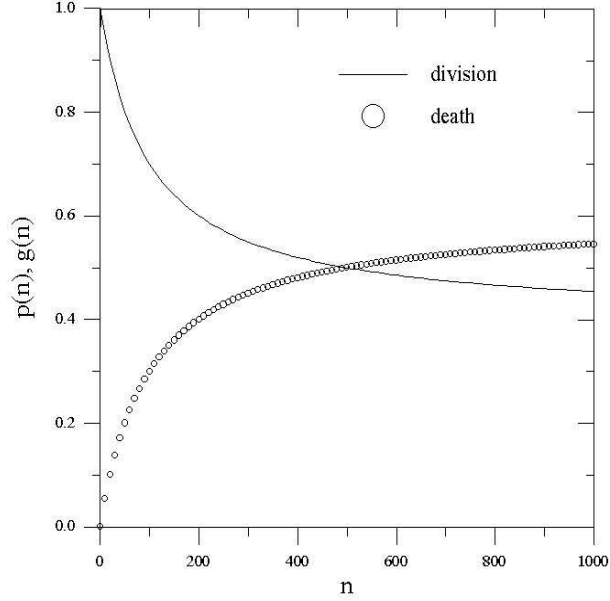}}
\end{center}
\caption{Division and death probabilities in the extend WB model.}
\label{pgcurves}
\end{figure}

Now, we introduce a new model by assuming that the division and death
probabilities depend on the total number of tumor cells $n$ through {\it
Michaelis-Mentem} functions \cite{biomod}: \begin{equation}
\label{probwbnew1}
g(n)=1-\frac{\alpha n}{\Gamma+n}
\end{equation}
and
\begin{equation}
\label{probwbnew2}
r(n)=\frac{\alpha n}{\Gamma+n}.
\end{equation}
Here $0<\alpha<1$ and $\Gamma>0$ are parameters that control the shape of the
curves.  These functional forms were used  because they are the simplest ones
which varies monotonically with $n$ and satisfy the  normalization condition
$g(n)+r(n)=1$. These curves are illustrated in figure \ref{pgcurves}.

\section{Analysis of WB models using stochastic methods}
\label{sc3}
The stochastic growth rules used in the WB models involve probabilities $g$ and
$r$ that  are explicitly time independent. In consequence, as is the case for
Markov chains, the chance of  these models generate a given pattern in a
certain time depends only on the probabilities associated to the configurations
at the previous time. So, the WB growth processes will be described within the
{\it probability transition equation} or {\it master equation} frameworks,
according to the discrete or continuous character of the time
\cite{vankampen,gardiner}.

The transition probability equation is written as:
\begin{equation}
\label{probtran}
P(n,t+1)=\sum_{m}T_{n,m}P(m,t),
\end{equation}
where $T_{m,n}$ are the elements of the {\it transition matrix}. $T_{m,n}$
provides the transition  probability from a state containing $m$ tumor cells to
a state with $n$ tumor cells in the next time. A continuous version of equation
(\ref{probtran}) is given by the master equation: \begin{equation}
\label{meqn}
\frac{d}{dt}P(n,t)=\sum_{m\neq n}\left\{ W_{n,m}P(m,t)-W_{m,n}P(n,t)\right \},
\end{equation}
where $W_{n,m}$ is interpreted as a probability by unit time or transition rate.

Initially,  the discrete time WB model will be considered. At every random
selection of an interfacial site to implement an action, the time is
incremented by an unity. In this way, the transition matrix is:
\begin{equation}
\label{tnm}
T_{n,m}= \left \{  
\begin{array}{ll}
g & \mbox{if $n=m+1$}\\
r & \mbox{if $n=m-1$}\\
0 & \mbox{if $|n-m|$}>1 
\end{array} \right. .
\end{equation}
This expression is valid  only for $n \ge 2$, since the WB model is a special
type of {\it one step process} \cite{vankampen} with an absorbent state at
$n=0$. Indeed, $T_{0,1}=r$ and $T_{1,0}=0$. Substituting the transition matrix
(\ref{tnm}) in the probability transition equation (\ref{probtran}) we have:
\begin{equation} \label{probtranwb}
\begin{array}{ll}
P(n,t+1)=gP(n-1,t)+rP(n+1,t) & \mbox{if $n\geq 2$}\\
P(n,t+1)=rP(n+1,t) & \mbox{if $n=0,1$}
\end{array} .
\end{equation}

In this section, we are interested  in quantities that depend only on the
number of tumor cells and not on the spatial distribution of theses cells on
the tissue. So, the time dependence of the average number of tumor cells
$\langle n(t)\rangle$ and its variance $\sigma (t)$, defined as:
\begin{equation} \label{mean}
\langle n(t) \rangle \equiv \sum_{n=0}^\infty nP(n,t)
\end{equation}
and
\begin{equation}
\label{sigma}
\sigma^2 (t) \equiv \langle n^2(t) \rangle -{\langle n(t) \rangle}^2,
\end{equation}
are calculated. Clearly,
\begin{equation}
\label{mean2}
\langle n^2(t) \rangle \equiv \sum_{n=0}^\infty n^2P(n,t).
\end{equation}
Using equation (\ref{probtranwb}) and iterating the expressions for $<n(t)>$ and $<n^2(t)>$ we find:
\begin{equation}
\label{solwbmean}
\langle n(t) \rangle=n(0)+\frac{\kappa-1}{\kappa+1}t
\end{equation}
and
\begin{equation}
\label{solwbsig}
\sigma^2 (t)=\left[ 1-\left(\frac{\kappa-1}{\kappa+1}\right)^2 \right]t.
\end{equation}
Equations (\ref{solwbmean}) and (\ref{solwbsig}) show that $\langle n(t)
\rangle$ decreases linearly with time if $\kappa<1$, and increases linearly  if
$\kappa>1$.  In turn, the variance increases with the square root of time.
Thus, for all $\kappa>1$ there is a non-vanishing probability of unlimited
tumor growth. But, if $\kappa=1$, $\langle n(t)\rangle$ is constant, the
variance increases with the square root of time and, therefore, independently
of the initial  population, the absorbent state $n=0$ will be reached.
Moreover, we have the maximum variance exactly at $\kappa=1$. These exact
results found for the discrete WB model are also valid for the continuous time
approach.

\section{The extended model}
\label{sc4}
In this section, the previously described generalized WB model (equations
(\ref{probwbnew1}) and (\ref{probwbnew2})), which includes the possibility of
growth limitation, is studied by a continuous approach based on the master
equation \cite{vankampen,gardiner}. Again, the time step is defined as the
birth or death of a single tumor cell.

\subsection{The macroscopic equation}
\label{ssc41}

First of all, we define the transfer matrix $W_{n,m}$:
\begin{equation}
W_{n,m}=\left\{ 
        \begin{array}{l}
        \displaystyle{g(m) \mbox{~if~} n=m+1}\\
        \displaystyle{r(m) \mbox{~if~} n=m-1}\\
	  \displaystyle{0 \mbox{~if~} |n-m|>1}
	  \end{array} 
        \right.
\end{equation}
or
\begin{equation}
\label{transmtxwbnew}
W_{n,m}=g(m)\delta (m-n-1)+r(m)\delta (m-n+1).
\end{equation}
The master equation is obtained substituting the expression 
(\ref{transmtxwbnew}) in (\ref{meqn}):
\begin{equation}
\label{meqnwbnew}
\frac{d}{dt}P(n,t)=g(n-1)P(n-1,t)+r(n+1)\times P(n+1,t)-P(n,t).
\end{equation}
Using (\ref{mean}), (\ref{mean2}) and (\ref{meqnwbnew}) we have:
\begin{equation}
\label{nmwbnew}
\frac{d\langle n(t) \rangle}{dt}=1-2\alpha \left\langle \frac{n(t)}{\Gamma+n(t)} 
\right\rangle
\end{equation}
and
\begin{equation}
\label{nmwbnew2}
\frac{d\langle n^2(t) \rangle}{dt}=1+2\left\langle \frac{n(t)[ \Gamma+(1-
2\alpha)n(t) ]}{\Gamma+n(t)} \right\rangle.
\end{equation}

Notice that (\ref{nmwbnew})  and (\ref{nmwbnew2}) are non-linear equations
which cannot be solved by analytical or numerical methods. Using a mean field
approximation, the macroscopic equation for (\ref{nmwbnew}) is obtained
replacing $n(t)$ by a deterministic function $ N(t)$. The corresponding
macroscopic equations for the first and second moments are:

\begin{equation}
\label{macroeq}
\frac{dN(t)}{dt}\equiv \zeta(t) =1-2\alpha \frac{N(t)}{\Gamma+N(t)}.
\end{equation}

\begin{equation}
\label{macroeq2}
\frac{dN^2_*(t)}{dt}=1+2\frac{N(t)[ \Gamma+(1-2\alpha)N(t) ]}{\Gamma+N(t)}.
\end{equation}
The star is used in order to distinguish between $N^2$ and $N^2_*$, the second moment.

\begin{figure}
\begin{center}
\resizebox{7cm}{7cm}{\includegraphics{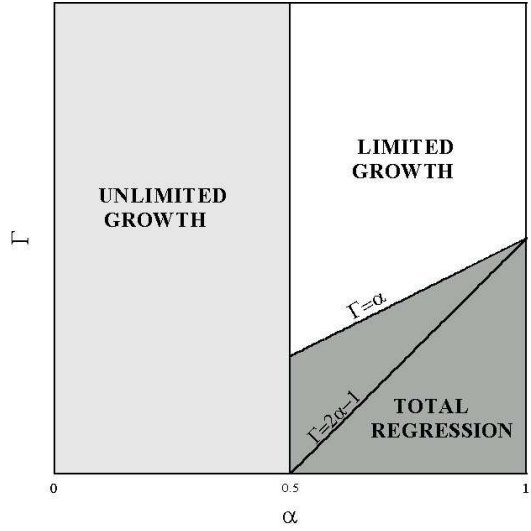}}
\end{center}
\caption{Mean field phase diagram  for the asymptotical value of the average
number of tumor cells. Three possible regimes are found: unrestricted growth,
limited growth and total regression of the tumor.} \label{diagram}
\end{figure}

As one can see, equation (\ref{macroeq})  exhibits three distinct asymptotic
behaviors: \begin{itemize}
\item $N(t)$ increases without limit if $\zeta (t)>0 ~~\forall~~t$ ;
\item $N(t)\rightarrow 0$ if $\zeta (t)<0 ~~\forall~~t$;
\item $N(t)$ goes to a non-vanishing  constant value if $\zeta (t)=0$ at a
certain time. \end{itemize}

\begin{figure}
\begin{center}
\resizebox{8cm}{8cm}{\includegraphics{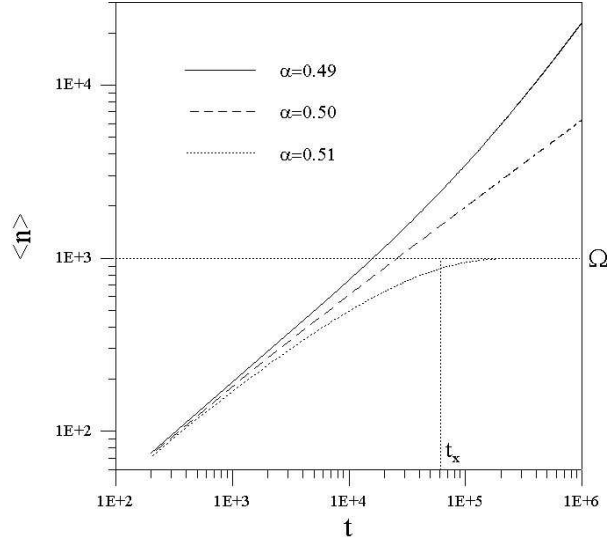}}
\end{center}
\caption{Numerical solutions of the macroscopic  equation for the extended WB
model around the critical point $\alpha=1/2$ and fixed parameter $\Gamma=20$.
Also, the saturation value and the crossover time evaluated for these
parameters are indicated.} \label{numsol}
\end{figure}

The phase diagram in the parameter space ($\Gamma$, $\alpha$)  is shown in
figure \ref{diagram}. The third behavior requires a stable solution
\cite{vankampen}, which always exists since the macroscopic equation is a first
order differential equation. This analysis reveals the existence of a
well-defined phase transition in $\alpha =1/2$, i. e., for $\alpha <1/2$ the
average number of tumor cells grows without limit, whereas for $\alpha > 1/2$
this number reaches a constant value $\Omega \ge 0$.

In the region of unrestricted growth ($\alpha<1/2$), $N(t)$ grows  linearly and
the variance increases as the square root of  the time, in agreement with
(\ref{macroeq}) and (\ref{macroeq2}) for very large $N$. Therefore, the model
behaves asymptotically as the original WB model. The transition line
$\alpha=1/2$ must be considered in separated. Fortunately, the equation
(\ref{macroeq})  has a closed solution for $\alpha=1/2$, $N_{\meio}(t)$, given
by: \begin{equation} \label{nmeio}
N_{\meio}(t)=\sqrt{(n(0)+\Gamma)^2+2\Gamma t}-\Gamma.
\end{equation}
Clearly, $N_{\meio}$ has the asymptotical behavior  $N_{\meio}(t)\cong
\sqrt{2\Gamma t}$. Substituting (\ref{nmeio}) in (\ref{macroeq2}), the
asymptotical macroscopic approximation to $N^2_*$ is obtained and, using this
result, we find $\sigma=\sqrt{t}$. However, numerical simulations suggest that
$\sigma\cong\sqrt{t/2}$, and this difference will be explained in the
subsection \ref{ssc43} by taking into account corrections to the macroscopic equation.

In the region of limited growth,  the solution $N_s(t)$ reaches a {\it
saturation value} $\Omega$ for long times. This value  can be find by taking
$\zeta (t)=0$: \begin{equation}
\label{satsize}
\Omega =\frac{\Gamma}{2\alpha-1}.
\end{equation}
Also, the {\it saturation crossover  time}, ($t_\times$),  is evaluated through
an expansion at long times of $N_s(t)$ around $\Omega$. Substituting
$N_s(t)=\Omega+\mu(t)$, where $|\mu(t)| \ll \Omega$, and expanding
(\ref{macroeq}) we have: \begin{equation}
\frac{d \mu}{dt}=-\frac{(2\alpha-1)^2}{2\alpha\Gamma}\mu+O(\mu^2).
\end{equation}
At first order in $\mu$ the solution  is an exponential decay $\mu(t)\sim
\exp(-t/t_\times)$. Thus, the saturation crossover time is given by:
\begin{equation}
t_\times=\frac{2\alpha\Gamma}{(2\alpha-1)^2}.
\end{equation}
In figure \ref{numsol} the numerical  integrations of (\ref{macroeq}) around
the critical parameter $\alpha=1/2$ and fixed  $\Gamma=20$ are shown.

The phase diagram exhibits two lines  in the complete regression region. The
line $\Gamma=2\alpha-1$ is the border of the region where $N(t)\rightarrow 0$,
according to the solution of the macroscopic equation. However, if the
fluctuations are larger than the saturation value $\Omega$, all tumor cells
die. Thus, the phase diagram shown in figure \ref{diagram} already take into
account the correction considering the variance,  as evaluated in next
subsection. The result is that the region of total regression correspond to
$\Gamma<\alpha$.

\begin{figure*}
\begin{center}
\resizebox{13.5cm}{13.5cm}{\includegraphics{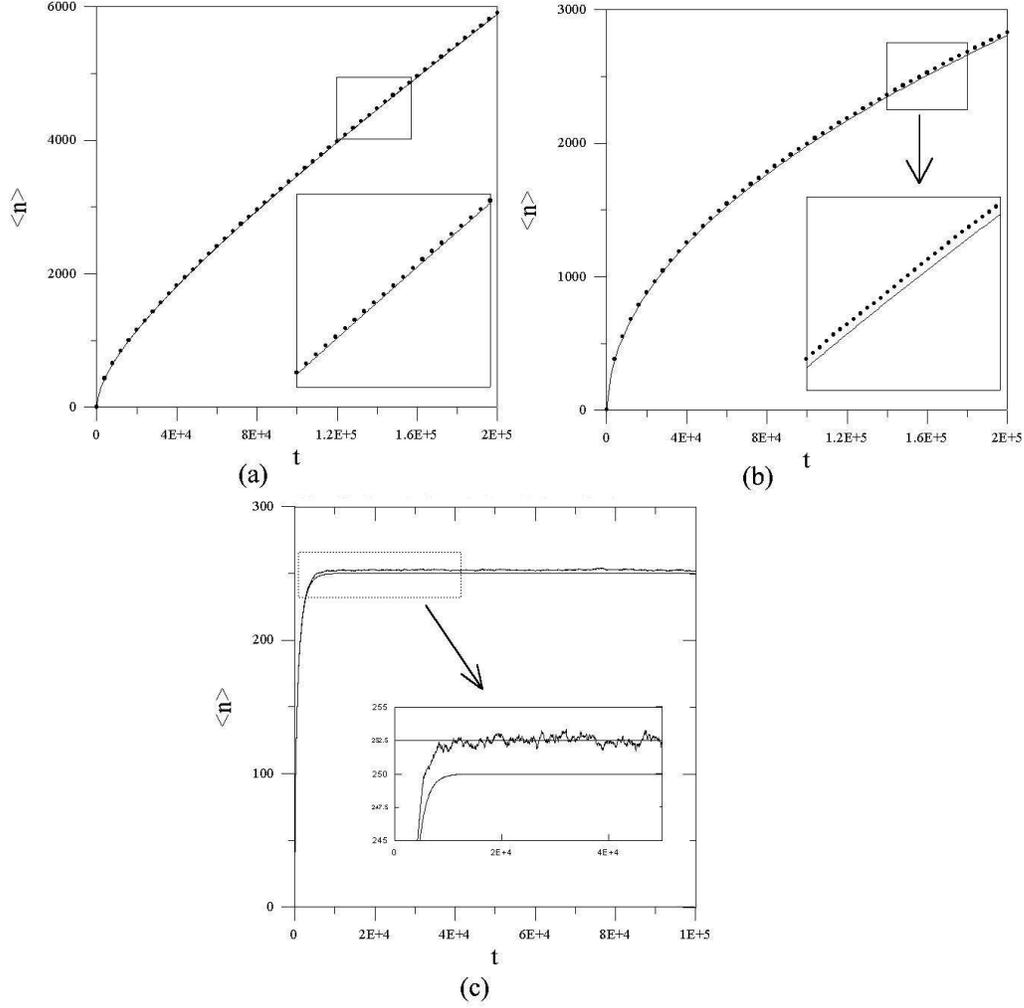}}
\end{center}
\caption{Comparison between  analytical and simulational results. The
continuous and smooth curves are obtained from numerical integrations of the
macroscopic equation whereas the other ones are given by simulations. 2000
samples were used in the simulations with the parameters (a) $\Gamma=20$ and
$\alpha=0.49$, (b)$\Gamma=20$ and $\alpha=1/2$ and (c) $\Gamma=50$ and
$\alpha=0.60$.} \label{compar} \end{figure*}

In order to test  the predictions of the macroscopic equation approximation,
comparisons with simulational results were done. It was observed a good
agreement between analytical and simulational results for $\alpha<1/2$.
However, for $\alpha\ge 1/2$,  small differences, that become meaningful when
the value of $\Gamma$ is not large, emerge. All these comparisons are shown in
figure \ref{compar}. As one can see, there is a fixed difference between the
numerical and simulated solutions in the limited growth region. This difference
can be found making an {\it expansion of the master equation} \cite{vankampen},
also called {\it $\Omega$ expansion}.

\subsection{The $\Omega$ expansion}
\label{ssc42}
The $\Omega$ expansion consists  in expand the master equation in powers of the
characteristic size of the system. In our case, this characteristic size is the
$\Omega$ in (\ref{satsize}), since it represents an upper bound to the size of
the tumor population. Indeed, this was the reason for denote the saturation
value as $\Omega$.

We assume that the probabilities $P_n$  have a sharp maximum around the
macroscopic solution with a width of order  $\Omega^{1/2}$. This hypothesis is
explicitly used in the change of variable: \begin{equation}
\label{changevar}
n=\Omega \phi(t)+\Omega^{1/2}\xi,
\end{equation} 
where $\xi$ is a new random variable and $\phi(t)$ is  a deterministic
function. Also, the expansion assumes that the rates $W(n|m)\equiv W_{n,m}$ can
be written in the form:
\begin{equation}
\label{canonical}
W_\Omega (n|m)=f(\Omega)\left\{ \Phi_0\left ( \frac{m}{\Omega};s
\right)+\Omega^{-1}\Phi_1 \left( \frac{m}{\Omega};s \right) \right. \nonumber
\\ \left. +\Omega^{-2}\Phi_2\left ( \frac{m}{\Omega};s \right) + \ldots
\right\},
\end{equation}
where $f$ and $\Phi_i$ are arbitrary functions and $s\equiv n-m$. The
technical details involved in the expansion can be found in \cite{vankampen}.

Introducing the notation used in \cite{vankampen}:
\begin{equation}
\varphi_{\nu,\lambda}(x)=\int s^\nu \Phi_\lambda (x;s)ds,
\end{equation}
the expansion up to orders of $\Omega^{1/2}$ and $\Omega^0$ results in the
following equations for $\phi$, $\langle \xi \rangle$ and $\langle \xi^2
\rangle$: \begin{equation} \label{phi}
\frac{d \phi}{d\tau}=\varphi_{1,0}(\phi),
\end{equation}
\begin{equation}
\label{xi}
\frac{\partial\langle\xi\rangle}{\partial\tau}=
\varphi_{1,0}^{\prime}(\phi)\langle\xi\rangle \end{equation} and
\begin{equation} \label{xi2}
\frac{\partial\langle\xi^2\rangle}{\partial\tau}=
2\varphi_{1,0}^{\prime}(\phi)\langle\xi^2\rangle+\varphi_{2,0} (\phi),
\end{equation}
with $\tau\equiv f(\Omega)/\Omega$.  (\ref{phi}) is the macroscopic equation
for the system divided by $\Omega$, (\ref{xi}) gives the correction for
$\langle n(t)\rangle$ in order of $\Omega^{1/2}$, and (\ref{xi2}) represents
the first approximation for the fluctuations around the mean. The initial
conditions for (\ref{xi}) and (\ref{xi2}) are
$\langle\xi(0)\rangle=\langle\xi^2(0)\rangle=0$, because
$P_n(0)=\delta(n-n_0)$, i.e., at $t=0$ the population is $n_0$.

For the extended model, the transition rates (\ref{transmtxwbnew}) can be
rewritten  as:
\begin{equation}
\label{canonicalwbnew}
W_\Omega(\rho;s)=\left( 1-\frac{\alpha\rho}{2\alpha-1+\rho}\right)\delta(s-1)
+ \left( \frac{\alpha\rho}{2\alpha-1+\rho}\right)\delta(s+1),
\end{equation}
where $\rho\equiv m/\Omega$. In agreement with the notation of
(\ref{canonical}), we have $f(\Omega)=1$, $\Phi_i=0\mbox{~if~}i\ne 0$, and
$\Phi_0$ is defined as the right hand side of (\ref{canonicalwbnew}).

Here, our goal is calculate the asymptotical corrections in time and the
differences shown in figure \ref{compar}. Solving (\ref{xi}) and (\ref{xi2})
using (\ref{canonicalwbnew}), we find that:
\begin{equation}
\label{xisol}
\langle\xi(\tau)\rangle\sim\exp\left(-\frac{2\alpha-1}{2\alpha}\tau\right)
\stackrel{\tau\rightarrow\infty}{\longrightarrow} 0
\end{equation}
and
\begin{equation}
\label{xi2sol}
\langle\xi^2(\tau)\rangle\stackrel{\tau\rightarrow\infty}
{\longrightarrow}\frac{\alpha}{2\alpha-1}.
\end{equation}
Therefore, the correction for $\langle n(\infty)\rangle$ given by
(\ref{xisol}) vanishes, because $n=\Omega \phi(t)+\Omega^{1/2}\xi(t)$. However,
(\ref{xi2sol}) shows that the first correction for the variance reaches a
constant value:
\begin{equation}
\label{varsol}
\sigma^2=\Omega\langle\xi^2(\tau)\rangle\stackrel{\tau\rightarrow\infty}
{\longrightarrow}\frac{\Gamma\alpha}{(2\alpha-1)^2}.
\end{equation}
In  figure \ref{compar}(c) the variance was  measured as
$\sigma_{measured}\cong 27.5$, whereas the analytical value calculated by
(\ref{varsol}) is $\sigma_{analytical}\cong 27.4$. Thus, there is an excellent
agreement between the analytical and simulational results for the variance.
Since the fluctuations reach a constant value which is smaller than the
saturation one, the tumor certainly does not disappear as in the original WB
model with $\kappa=1$.

The  difference between the saturation values obtained by the simulations and
calculated through the macroscopic equation can not be explained taking into
account only terms of order $\Omega^0$. The first non-vanishing correction in
$\langle\xi\rangle$ involves terms of order $\Omega^{-1/2}$\cite{vankampen}:
\begin{equation}
\langle\xi\rangle=\frac{\Omega^{-1/2}}{2(2\alpha-1)}.
\end{equation}
Consequently, the first correction in $\langle n\rangle$ is:
\begin{equation}
\label{delta}
\Delta=\frac{1}{2(2\alpha-1)}.
\end{equation}

Using (\ref{delta}), the calculated correction is $\Delta_{analytical}=2.5$
and the measured value in figure \ref{compar}(c) is $\Delta_{measured}\cong
2.51$, which are in excellent agreement. Several other values for the model
parameters were tested and a very good agreement between simulational and
analytical results was found.

\subsection{Asymptotical corrections for $\alpha=1/2$}
\label{ssc43}

The $\Omega$ expansion cannot be applied for systems without a characteristic
size. This is the case for equation (\ref{nmeio}) which, for $\alpha=1/2$,
gives a solution that grows without limit and, consequently, one can not define
the characteristic size of the tumor. However, it is possible to do an
expansion around the macroscopic solution (\ref{nmeio}). Taking
$n(t)=N_{\meio}(t)+\varepsilon(t)$, where $\varepsilon$ is the new random
variable satisfying $|\varepsilon(t)|/N_{\meio}(t)\ll 1$, and conserving terms
up to second order of $\varepsilon/N_{\meio}$, we find the following equation
for $\varepsilon(t)$:
\begin{equation}
\label{eqvarepsilon}
\frac{d<\varepsilon>}{dt}=-\Gamma\frac{<\varepsilon>}{g^2(t)}+\Gamma\frac{<\varepsilon^2>}{g^3(t)}
+O(\varepsilon^3),
\end{equation}
where $g(t)=\sqrt{n(0)+2\Gamma t}$. In order to solve
(\ref{eqvarepsilon}), it is necessary to know the relationship between the
first and the second moments of $\varepsilon$. Equation (\ref{macroeq2}) gives
$N^2_*(t)=N_{\meio}^2 (t)+t$ and the xpansion of
$n^2=(N_{\meio}+\varepsilon)^2$  with the approximation $<n^2>\approx N^2_*$
results in $<\varepsilon ^2>\approx t-2N_\meio<\varepsilon>$. Substituting this
approximated relation for $<\varepsilon ^2>$ in equation (\ref{eqvarepsilon}),
and taking the asymptotical limit for $t$, we obtain: \begin{equation}
\label{eqvarepsilon2}
\frac{d<\varepsilon>}{dt}+\frac{3}{2t}<\varepsilon>-\frac{1}{2\sqrt{2\Gamma
t}}=0. \end{equation} Its exact solution is: \begin{equation}
\label{varepsilon} <\varepsilon(t)>=\sqrt{\frac{t}{32\Gamma}}. \end{equation}
Considering the correction given by (\ref{varepsilon}), the analytical and
simulational curves in figure \ref{compar}(b) become indistinguishable. Now,
one can use the result (\ref{varepsilon}) to find the correction to the
variance. It is possible to show that the asymptotical value of the variance,
consistent with the approximations used above, is given by: \begin{equation}
\label{varmeio} \sigma^2=\frac{t}{2}\left(1-\frac{1}{16\Gamma} \right)
\end{equation} Equation (\ref{varmeio}) is able to explain the factor $1/2$
present at the simulations shown in figure \ref{compar}(b).  Again, the
simulational and analytical results are in very good agreement for a large
number of parameter sets.

\begin{figure*}
\begin{center}
\resizebox{13.5cm}{13.5cm}{\includegraphics{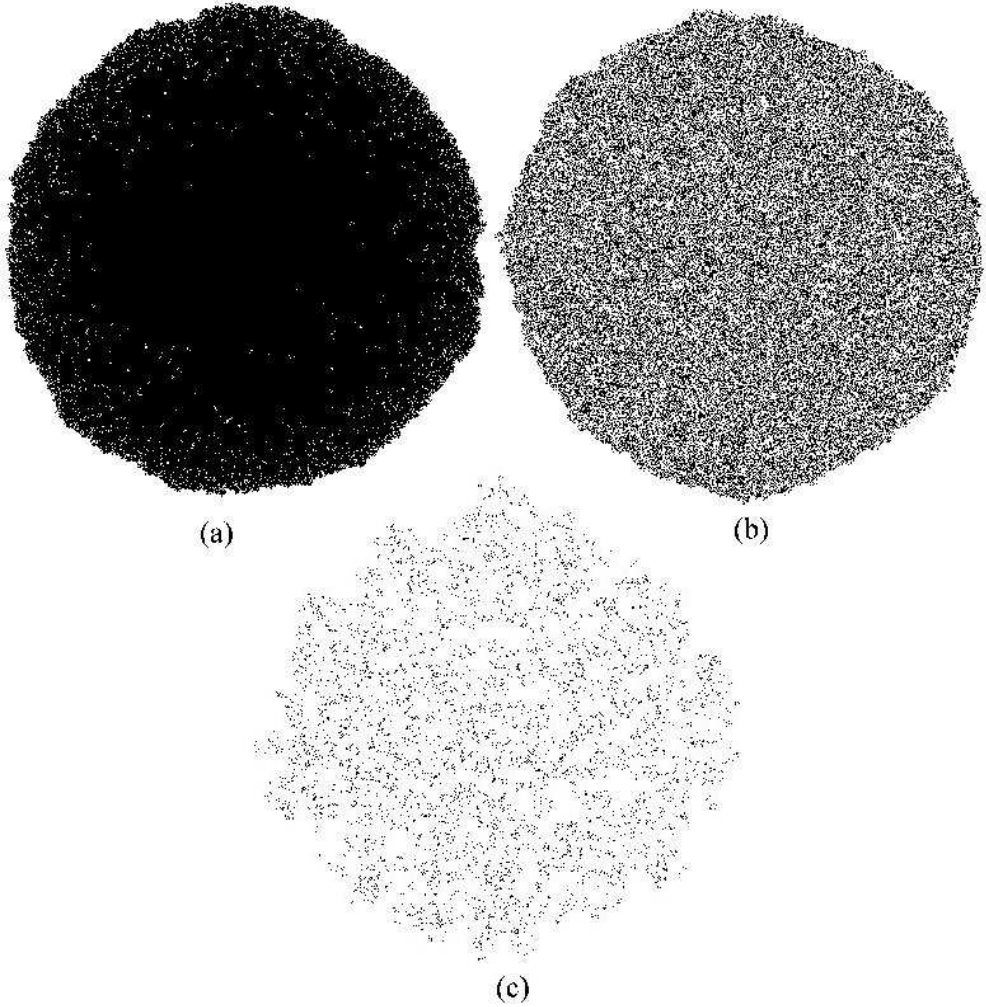}}
\end{center}
\caption{Typical patterns generated by the extended WB model: (a) compact, (b) connected and (c) disconnected. For these patterns $\Gamma=1000$ were used. $\alpha=0.45$, $0.50$ and $0.51$ was used in (a), (b) and (c), respectively. The simulations were done in lattices with $1200 \times 1200$ sites and stopped if a tumor cell reaches the border of the lattice.}
\label{wbpatterns}
\end{figure*}

\section{Geometrical patterns}
\label{sc5}

The previous sections were  dedicated to the analytical study of the original
WB model and its extended version. Now, in this section  the simulational
results focusing the geometrical properties of the patterns generated by the
extended WB model are presented. The patterns associated to the original WB
model are spherical and compact with a rough surface. For the extended model,
the patterns exhibit three distinct morphologies: compact with a rough border,
connected with internal holes and disconnected with cells isolated from each
other. These morphologies are shown in figure \ref{wbpatterns}. In the region
of unlimited growth, i. e., $\alpha<1/2$, the patterns become compact. This
compaction occurs because the division probability is always higher than the
death probability and, as a consequence, all the internal empty sites will be
occupied at sufficiently long time. The disconnected patterns appear in the
region of limited growth ($\alpha>1/2$). The reason is that the division and
death probabilities reach the same value $p=1/2$, and the cells behave like
non-directed random walkers. Finally, the connected patterns are generated just
on the transition line $\alpha=1/2$. The difference between connected and
disconnected patterns is that the former does not has a percolation cluster of
empty sites inside the pattern, whereas in the last this cluster is found.

These patterns were characterized by its gyration radius ($R_g$) and the
number of interface tumor cells ($S$). The gyration radius is defined by:
\begin{equation} \label{rg}
R_g=\sum_{i=1}^{n}(\vec{r_i}-\vec{r}_{cm})^2,
\end{equation}
where the sum extends over all the tumor cells and $\vec{r}_{cm}$ is the mass center
of the pattern. In the original WB model, $R_g$ and $S$ scale asymptotically
with the square root of the number of tumor cells \cite{scfj}. In turn, only
the compact patterns ($\alpha<0.5$) of the extended model show the same
asymptotical behaviour for $R_g$ and $S$. For $\alpha \ge 0.5$ the scaling laws
change and are dependent on the $\Gamma$ parameter. Indeed,  $S\sim n$ and
$R_g\sim n^\nu$, with $\nu(\Gamma)>1/2$, indicating that the patterns are
fractals with dimensions $d_f=1/\nu$ \cite{vicsek}. Notice that, in the limited
growth regime, $n$ is bounded and the power law is defined before the growth
saturation. However, since the cells become progessively more distant from each
other, $R_g$ grows continuously with the time as a power law. The numerical
results are summarized in table \ref{tabela}.

The transition from compact to disconnected patterns was studied through the  density $\rho$ of
internal holes in the patterns. $\rho$ was defined as the ratio between the number of empty and
occupied sites enclosed by a circle of radius $R_g$. This definition was used in order to discard
 the tumor border, where the compaction never occurs. For $\alpha<1/2$, $\rho$ decreases, vanishing
at sufficient long time. In contrast, for $\alpha \ge 1/2$, $\rho$ increases asymptocally as a power
law $\rho\sim t^\gamma$.

In order to test  if the exponent $\gamma$ varies with the parameter $\Gamma$,
simulations were done using three distincts $\Gamma$ values
($10^2$,$10^3$,$10^4$). For $\Gamma=10^4$ the density of internal holes is lower
than that for $\Gamma=10^3$, but the corresponding power law exponent is
higher. Since for both $\Gamma=10^2$ and $\Gamma=10^3$ the simulations provide
essentially the same values for  $\gamma$, we suppose that this exponent is
independent of the parameter $\Gamma$. Thus, we believe that the curve obtained
for $\Gamma=10^4$ might be a transient behavior.

\begin{table}[t]
\caption{Summary of the  exponents found for the extended WB model. The
exponents $\nu$, $\sigma$, and $\gamma$ are defined by $R_g\sim n^\nu$,$S\sim
n^\sigma$ and $\rho\sim t^\gamma$.}

\begin{center}
\begin{tabular}{|c|c|c|c|c|c|c|c|c|} \hline 
\label{tabela}
           & \multicolumn{2}{c|}{$\alpha<1/2$} & \multicolumn{3}{c|}{$\alpha=1/2$} & \multicolumn{3}{c|}{$\alpha=0.51$} \\ \hline
$\Gamma$   & $\nu$   & $\sigma$ &  $\nu$  & $\sigma$& $\gamma$  &  $\nu$  & $\sigma$& $\gamma$   \\ \hline 
$10^3$     & $0.50$  &  $0.50$  &  0.64   &   ~1~   &  0.18     &   0.62  &    ~1~  &  0.64   \\ \hline
$10^3$     & $0.50$  &  $0.50$  &  0.60   &   ~1~   &  0.18     &   0.60  &    ~1~  &  0.64   \\ \hline
$10^4$     & $0.50$  &  $0.50$  &  0.54   &   ~1~   &  0.25     &   0.52  &    ~1~  &  0.63   \\ \hline 
\end{tabular}
\end{center}
\end{table}

The patterns observed in the  extended model are very similar to the
morphologies generated by a growth model for primary cancer recently proposed
\cite{scfj2}. This model considers a complex interaction network among tumor
cells mediated by growth factors and, in contrast to the present model,
involves a large number of parameters. Our results are more robust than those
observed in \cite{scfj2}, since the two parameters used in the extended WB
model are not associated to the properties of the tumor microenvironment.
Indeed, they are related to macroscopic quantities such as the saturation size
of the tumor. However, the extended  WB model has a limitation. Every compact
pattern has a unlimited growth and the disconnected patterns always cease their
populational growth. The reason is that the growth only occurs on the interface
between tumor and normal tissues, whereas in real tumors the internal cells
also divide. This limitation does not compromise the analytical results since
the spatial distribution of the tumor cells in the tissue is not considered.

\section{Conclusions}
\label{sc6}

In this work we propose an extended  version of the Williams and Bjerknes (WB)
model initially used to describe the tumor growth. In this extended model, the
division and death probabilities correspond to monotonically  increasing and
decreasing functions of total number of tumor cells $n$, respectively. The
average values of $n$ for the original and the extended WB models as well as
their variances were exactly calculated using stochastic processes methods. The
original model reveals two possibilities for $<n>$: unlimited growth or total
regression of the tumor. However, real tumors exhibit three possible behaviors
which are observed also in the extended WB model: the ones previously described
and a quiescent state where the tumor size remains constant for a long time.
The differences observed between simulated and mean-field results for the
extended model were calculated using expansions of the involved equations. In
addition, the geometry of the growth patterns generated by the extended WB
model were analysed. Three distinct morphologies have been observed: compact,
connected and disconnected. All the patterns were characterized by its gyration
radius, $R_g$, and number of tumor cells on the interface, $S$. For the compact
patterns, $R_g\sim \sqrt{n}$ and $S\sim\sqrt{n}$. For the connected and
disconnected patterns $R_g\sim n^\nu$ and $S\sim n$, with $\nu>1/2$, indicating
that these patterns are fractals.

{\bf Acknowledgements}\\ \\
We wish to thank Professor Ronald Dickman (Derpartamento de F\'{\i}sica, UFMG,
Brazil) and Professor M. L. Martins (Departamento de F\'{\i}sica, UFV, Brazil)
for enlightening discussions and important suggestions to improve the
manuscript. This work was supported by the CNPq-Brazilian agency. \\ \\


\end{document}